# Distinct photoluminescence in multilayered van der Waals heterostructures of $MoS_2/WS_2/ReS_2$ and BN

U. Bhat,[1] R. Singh,[1] B. Vishal,[1] A. Sharma,[1] H. Sharona,[1] R. Sahu,[1] and R. Datta[1,*]

[1]*International Centre for Materials Science, Chemistry and Physics of Materials Unit, Jawaharlal Nehru Centre for Advanced Scientific Research, Bangalore 560064, India.*

van der Waals heterostructures of ($TMD_{L=1}/BN_{L=1-4}/TMD_{L=1}/BN_{L=1-4}$), [TMD = $MoS_2$, $WS_2$, and $ReS_2$] are grown on *c*-plane sapphire substrate by pulsed laser deposition under slow kinetic condition. The heterostructure systems show strong emission around 2.3 eV and subsidiary peaks around 2.8, 1.9, 1.7 and 1.5 eV. BN and TMDs forms type-I heterojunction and the emission peaks observed are explained in terms of various band to band recombination processes and considering relative orientation of Brillouin Zones. The emission peak around 2.3eV is promising for solar and photovoltaic application. The observation is almost similar for three different heterostructure systems.

*Corresponding author e-mail: ranjan@jncasr.ac.in

The journey of 2D materials began after the discovery of graphene (G) with the hope of shrinking device dimension following Moore's law.[1] However, lack of a band gap in G shifted the attention towards alternative layered 2D materials, e.g., $MoS_2$, $WS_2$, $ReS_2$ etc. which belong to the family known as transition metal dichalcogenides (TMDs).[2] Many unique properties based on graphene and TMDs show promising novel device applications.[3-6] Vertical or in-plane arrangement of layers between various van der Waals (vdW) compounds, called van der Waals heterostructures, and superlattices have emerged as a new class of material system with immense possibility of tuning the existing properties of individual layer and exploring new physical phenomena.[7-10] The motivation for research on van der Waals heterostructures got momentum after the demonstration of significant increase in the mobility of graphene on *h*-BN substrate.[11] *h*-BN substrate also increased the mobility of $MoS_2$, $WSe_2$ significantly.[12] Moreover, insulation of graphene between *h*-BN sheet, mimicking the layered superconductor structure with greater scope for varying the materials parameters, exploration of fundamental physical phenomena e.g., Hofstadter butterfly effect in G/*h*-BN heterostructure, secondary Dirac points etc. have already been demonstrated.[13-15]

Addition of TMDs, particularly $MoS_2$, $WS_2$, and $ReS_2$ in the library of useful 2D materials broadened the horizons of van der Waals heterostructures significantly. Integration of TMDs with *h*-BN and G, offers the possibility to investigate the fundamental physical phenomena e.g., Bose Einstein Condensation (BEC) of excitons and to realize various novel heterostructure devices like field effect tunneling transistors based on G/*h*-BN & G/$WS_2$ vertical stacking[16,17] resonant tunneling diodes ($MoS_2$/BN), light emitting diodes, devices based on coupling between spin and valley degrees of freedom, strong light mater interaction and distinct excitonic behavior.[18,19]

Thinnest *p-n* junction based on $MoS_2$-$WSe_2$ vertical heterostructure has been demonstrated.[20] Among various device options, *p-n* junction forming type-II heterostructure can convert photon into $e^-$-$h^+$ pairs and extremely useful for solar photovoltaic and photodetector applications. Type-II heterostructures based on various combinations of TMDs e.g., tuning the optical emission in $MoS_2$/$WSe_2$, $MoS_2$/$WS_2$, fast interlayer energy transfer in $MoSe_2$/$WS_2$ for optical amplification and energy harvesting have been demonstrated.[21-23] On the other hand, it is the type-I heterostructure which was reported to form between $MoS_2$-$ReS_2$ with long charge transfer time of the order of 1 ps.[24] The type-I heterostructure is suitable for optoelectronic devices like light emitting diodes (LEDs) and laser diodes (LDs). There are reports on the heterostructures between *h*-BN and TMDs.[21,25-27] Electronic decoupling occurs between $MoS_2$/$WSe_2$ heterostructure with more than two layers of *h*-BN in-between TMDs. Theoretically, type-I heterostructure has been reported to form between *h*-BN and $MoS_2$. However, experimental verification on the formation of type-I heterostructure between TMDs and *h*-BN is not reported so far.

Nevertheless, arranging such vdW 2D-materials on top of each other or zipping on the sides is not straightforward. Early attempts involved a stacking procedure, first by isolating individual layers on a thin transparent polymer film and sticking them face to face.[28] The steps are lengthy and prone to contamination at the interface between vdW layers. Self-organization of various 2D materials from the suspensions was also reported. However, from the practical point of view, large area wafer scale thin film crystal growth methods like chemical vapor deposition (CVD), pulsed laser deposition (PLD), atomic layer deposition (ALD) are the viable options. Micrometer sized large crystals and their heterostructures of $MoS_2$/*h*-BN, have been grown by CVD.[29-31] PLD has also been employed under slow kinetic condition to grow epitaxial films of $MoS_2$ and $WS_2$ over large area on *c*-plane sapphire substrate with control over layer numbers with the retention of substrate induced strain and resulting modification in the band gap.[32,33] The success of the slow growth kinetics of PLD has been extended further to deposit different multilayered heterostructure stacks of $MoS_2$/$WS_2$/$ReS_2$ and BN on *c*-plane sapphire substrate. Such stacks are important not only for energy and optoelectronic applications but also to form coupled quantum wells (CQWs) heterostructure. CQWs provides a system to explore indirect excitons (IXs) which may form between the electrons and holes located at different layers separated by a distance.[34,35] IXs may form quantum degenerate Bose gas.

In the present report, formation of large area multilayered type-I heterostructures between three different TMDs and *h*-BN is demonstrated. Transmission electron microscopy imaging confirms the formation of such heterostructure systems. The heterostructure system shows a strong emission peak around 2.3 eV in the PL spectra irrespective of the TMD/BN system and is close to the most intense peak of the solar spectrum. This may prove extremely efficient as photovoltaic devices in suitable combination with type-II heterojunctions. The proposal is viable as BN can be doped both *p* and *n* type and $MoS_2$/$WS_2$/$ReS_2$ can easily be doped *n* type, therefore, charge extraction is possible. Subsidiary peaks around 2.8, 1.9, 1.7 and 1.5 eV are also observed with reduced intensity. The emission peaks observed are explained in terms of various band to band recombination processes and considering relative orientation of Brillouin Zones of TMDs and BN.

Heterostructures of $MoS_2$ (or $WS_2$ or $ReS_2$)/*h*-BN/$MoS_2$ (or $WS_2$ or $ReS_2$) were grown by PLD technique under slow kinetic condition. $MoS_2$, $WS_2$, and *h*-BN target pellets were prepared from powders obtained from Sigma Aldrich (99.9% purity) by first cold pressing and then sintering at 500 °C for 5 hours in a vacuum chamber (~$10^{-5}$ Torr). Sintering in the vacuum chamber prevents oxidation of compounds as well as re-deposition of vapor species back on the pellet surface unlike sintering performed in a sealed quartz tube. The $ReS_2$ target pellets were prepared following a procedure already reported earlier.[36] The epitaxial growth method of all the heterostructure was followed from the procedure based on PLD already developed for epitaxial growth of $MoS_2$ and $WS_2$ thin films on *c*-plane sapphire substrate.[33] However, number of pulses required to control layer numbers were further optimized for the growth of different TMDs and BN layers on top of each other. Kindly note that the PLD method was not utilized before to grow such van der Waals heterostructures, which was done either by physically stacking layers on top of each other or by CVD process to grow only two different layers.[28,29-31] Briefly, present growth method was a two-step process involving first deposition of a nucleation layer at 400°C and then final growth at 800°C. The laser ablation frequency was 1 Hz to ensure smooth film of such vdW layers with the underlying substrate and layers on top of each other. The number of layers of TMDs was one and in-between, one to four layers of *h*-BN were deposited. The top layer was encapsulated with *h*-BN layer to minimize external interaction.

Cross sectional transmission electron microscopy (TEM) specimen preparation was carried out by first mechanical polishing and then Ar ion milling to perforation to generate large electron transparent thin area. Special care was taken during sample preparation to maintain the vdW stacks intact which is otherwise extremely difficult as they delaminate with slight shear during joining pieces together. All the microscopy investigation was carried out in an 80-300 kV double aberration corrected transmission electron microscope.

Raman spectra were recorded using a custom-built Raman spectrometer using a 514.5 nm laser excitation and grating of 1800 lines/mm at room temperature. The laser power at the sample was approximately 1 mW. Micro PL measurement was performed with LabRAM HR (Horiba Jobin YVON) of spot size 1 micrometer with CCD detector and 1800 lines/mm grating.

Fig. 1(a) is the schematic representation of the three different vdW-heterostructures [($TMD_{L=1}/BN_{L=1-4}/TMD_{L=1}/BN_{L=1-4}$), TMD = $MoS_2$, $WS_2$, and $ReS_2$] grown by PLD with varying number of BN layers. The same number of BN layers in between TMDs and on top of the TMD layer is grown to form the heterostructure. Fig. 1 (b) &(c) are the example HRTEM images of such heterostructure with two separate layers of $MoS_2$ and $WS_2$ and four BN layers in-between. It is extremely difficult to prepare cross sectional TEM specimens with few layers of layered materials on the substrate as they are found to be delaminated unless extreme care is taken.[33] As already mentioned, the number of layers of different materials can be controlled with the laser ablation frequency under slow kinetic growth condition and example HRTEM images of scaled up version of such heterostructures are given in the supplementary document [Fig. S1].

Raman spectra of BN layers deposited on top of three different TMDs are displayed in Fig. 2 (a)-(d). It is found that the crystal structures of BN grown on top of TMDs are different depending on the type of the TMD compound; *h*-BN on $MoS_2$, mixture of *c*-BN and *w*-BN on $WS_2$ and $ReS_2$. The Raman spectra of powder *h*-BN, *c*-BN and *w*-BN are provided for the reference purpose [Fig. S2].[37] The peak around 1370 $cm^{-1}$ corresponds to the *h*-BN structure. For *c*-BN there are two distinct peaks around 1057 and 1309 $cm^{-1}$ and many broad peaks appears in case of *w*-BN structure. In the present heterostructure samples, the BN films grown on $MoS_2$ shows peak around 1365 $cm^-$ confirming the hexagonal structure [Fig. 2 (a)]. The Raman peaks of BN grown on $WS_2$ and $ReS_2$ matches closely with the *c*-BN and *w*-BN structures [Fig. 2 (b)-(d)]. However, almost similar PL spectra is obtained for three different TMD heterostructure systems, irrespective of the structure of BN and this may be because of the electronic structure in terms of band gap remains almost the same for different structural configurations During this study, it is also found that BN grows with wurtzite structure directly on the sapphire and details of the findings will be reported as part of a separate paper [Fig. S3]. The results show that the various crystal forms of BN can be controlled with the choice of different TMD materials as a template. The results are interesting as it is known that the synthesis of *c*-BN and *w*-BN requires extreme temperature and pressure conditions like diamond.[38,39]

The photoluminescence spectra at room temperature for three different heterostructures with varying number of BN layers show five distinct peaks labeled as A, B, C, D, and E around 435 (2.85), 532 (2.33), 595 (2.08), 720 (1.72) and 845 (1.47) nm (eV), respectively for $MoS_2$/BN heterostructure system. For other two systems, almost similar peak positions are observed [Fig. 3 (a)-(c)]. The various peak positions and full width at half maxima (FWHM) for different heterostructure systems are given in Table I. Moreover, peak positions and spectral pattern are almost similar irrespective of the number and crystal structure of BN layers deposited in between the TMDs. However, relative peak intensities are found to be slightly different with different number of BN layers.

Among five different peaks, peak B is the most intense, followed by peak A, C, E and D [Fig. 3(a)]. This is consistent between various heterostructure systems. To understand the origin of different peaks, electronic structure calculations based on the MoS$_2$/BN heterostructure is considered.[26,27] The calculated band structure and density of states (DOS) of MoS$_2$/BN heterostructure are given in the supplementary document [Fig. S4]. The dominant transitions are marked between various valence band to conduction band states and a schematic transitions levels derived from this is shown in Fig. 4. The calculation predicted type-I heterojunction between MoS$_2$ and *h*-BN system and similar heterojunction can be expected in case of WS$_2$ and ReS$_2$ based systems as well. The valence band maxima of MoS$_2$ in the partial density of states (PDOS) is higher than that of *h*-BN by about 0.05 eV, which is significantly smaller compared to the difference in conduction band minima levels by about 3.1/3.55 eV predicted by PBE+D2/HSE12 functional. In the schematic energy band diagram [Fig. 4] the band gap of *h*-BN and MoS$_2$ are 4.82/6.05 and 1.75/2.5 eV, respectively, which are estimated by PBE+D2/HSE functional.[27]

Monolayer MoS$_2$ has a direct band gap of 1.67 eV at K point and indirect band gap of 1.78 eV at $\Gamma \rightarrow K$ point.[40] Whereas, theoretical calculation of *h*-BN shows direct and indirect transitions at 4.69 eV and 5.50 eV points.[41] For the MoS$_2$/*h*-BN heterostructure, three possible transitions have been identified from the calculated band structure, these are: 2.857 $\Gamma \rightarrow K$, 2.543 eV at $M \rightarrow K$, and 1.57 eV at $K \rightarrow K$ points [Fig. S4].[26] The first two transitions at 2.5 and 2.8 eV are also marked in the DOS of the compound MoS$_2$/*h*-BN system. Now from the various transitions as marked in the combined DOS of the MoS$_2$/*h*-BN heterostructure system, transitions at 2.8 and 2.3 eVs can be assigned to the peak A and B of the PL spectra, respectively. Peak C and D at 2.1 and 1.72 eV can be assigned to the MoS$_2$ A and B excitonic peaks. Slight difference in peak values between experiment and theory can be attributed due to inherent errors involved with calculation methods, unknown effects like impurities, charges present in between the layers etc. The low energy peak around 1.47 eV could be because of defects or edge state levels for the combined system. For $MoS_2$ and $WS_2$ defect related spectral feature 0.1 eV below the excitonic peak and enhanced photoluminescence from defects have been reported and for $ReS_2$ no changes in photoluminescence due to defect is reported.[42-44] In the present heterostructure systems planar faults due to discontinuous Mo atoms are commonly observed and could also be responsible for the emission peak around 1.47 eV [Fig. S5].

The room temperature broadening of PL peaks of any semiconductor is important as this directly affects the performance of LEDs, LDs and other devices. The FWHM of all the peaks is broader compared to the typical broadening observed for the emission from monolayer $MoS_2$, $WS_2$ and $ReS_2$ [Table 1]. The room temperature broadening of PL peaks of the quantum wells semiconductors may be due to following possibilities; strong exciton-phonon interaction at room temperature, increase in thermal population at higher electron and hole energy levels, scattering by longitudinal-optical (LO) phonons, average well width fluctuations and associated quantum coupling effect, scattering due to impurities, dopants and longitudinal acoustic (LA) phonons.[45] The variation in thickness of BN barrier layers and $MoS_2$ layers can be expected in the PLD grown TMD/BN heterostructure system [Fig. S1]. Two vastly different atoms and crystal lattices of TMDs and BN may contribute to the broadening of the PL peaks.[24] Moreover, encapsulation of MoS2 by *h*-BN was predicted to increase the dissociation rate of excitons which may also contribute to the broadening of the emission peaks.[46] Therefore, the fluctuation of well and barrier width, longitudinal phonon scattering at room temperature and momentum vector mismatch may have contributed to the observed PL peak broadening.

Now the most intense peak at 2.31 eV in case of $MoS_2$/$h$-BN system and similar peaks for other two systems can be understood as follows; as already mentioned, the difference between the valence band edges of TMDs and BN is very small i.e. 0.05 eV. Therefore, the valence band edge of the compound TMD/BN system have contributions from both TMDs and BN. However, with increasing number of BN layers in between and above the TMDs layer, the valence band is expected to be dominated by the BN DOS. Though, the theory calculation suggests that the transition at 2.3 at M $\rightarrow$ K is an indirect transition and should not be most intense for the compound system where at K $\rightarrow$ K (1.57 eV) direct transition exists. However, the calculation was carried out with a specific stacking of $h$-BN on $MoS_2$ where the respective Brillouin zones have same orientation. In fact, the energy difference is negligible between various stacking arrangement between $h$-BN and $MoS_2$ because of the weak quantum mechanical interaction between the two systems. The band structure in terms of relative position of various VBM and CBM between $h$-BN and $MoS_2$ can be different in the momentum space depending on the relative rotation between the Brillouin zones. The transitions around 2.3 eV can be made strong if the VBM of $h$-BN at M point align with the CBM of $MoS_2$ at K point. The corresponding Brillouin zones for the system is shown in Fig. 5 (a) & (b) with the resulting band structure in Fig. 5 (c) & (d). As we could not verify the relative orientations between the $h$-BN and TMDs in our experimental heterostructure systems, therefore we anticipate based on the PL intensity peak and theory calculation that the orientation between the two crystal layers is according to Fig. 5 (d). This can explain strong emission peaks around 2.3 eV in the PL spectra. The transition at 2.8 eV is between the VBM at Γ point of $h$-BN to the CBM at K point of $MoS_2$. In addition to this the effect of electric dipole due defects can tilt the balance from indirect to direct transitions.[26,47]

In conclusion, large area growth of heterostructures of three different TMD and BN layers by PLD is demonstrated. Raman spectra shows BN can form with different crystal structure depending on the type of TMD as template layer. PL spectra shows strong emission around 2.3 eV and may be a good candidate for solar photovoltaic application.

See supplementary material for the additional TEM images for scaled up version of heterostructure, Raman spectra, schematic band structures and defects in the epitaxial TMDs film.

The authors at JNCASR sincerely acknowledge Prof. CNR Rao for advanced microscopy facility and ICMS for the funding. R. Datta also thank Dr. Somak Mitra and Iman Roqan at KAUST for some of the PL measurements.

**Table 1.** Experimental peak positions and FWHM of various peaks of different heterostructure systems.

|  | A | | B | | C | | D | | E | |
|---|---|---|---|---|---|---|---|---|---|---|
| TMDs | Peak (nm) | FWHM (nm) | Peak (nm) | FWHM (nm) | Peak (nm) | FWHM (nm) | Peak (nm) | FWHM (nm) | Peak (nm) | FWHM (nm) |
| $MoS_2$ | 435 | 61.25 | 532 | 77.69 | 595 | 123.22 | 720 | 35.14 | 845 | 56.18 |
| $WS_2$ | 432 | 58.2 | 545 | 84.09 | 643 | 103.60 | 717 | 41.44 | 840 | 54.52 |
| $ReS_2$ | 426 | 57.78 | 544 | 98.93 | 645 | 91.23 | 712 | 52.36 | 844 | 50.42 |

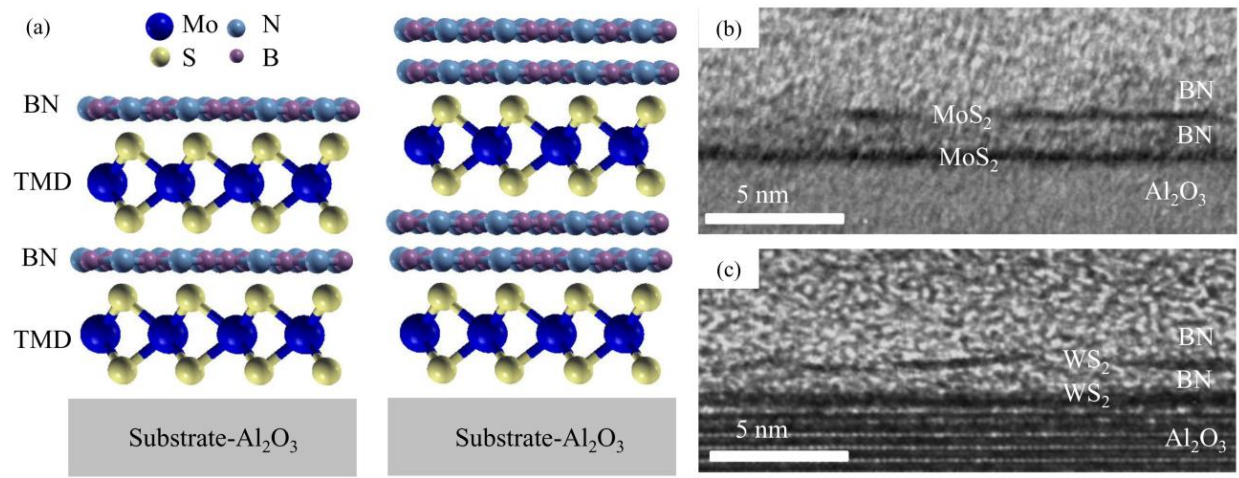

**Figure 1.** (a) Schematic of TMDs and *h*-BN vdW heterostructures with varying number of *h*-BN layers in between. Example cross sectional TEM image of (b) $MoS_2$ L=1/BN L=4/$MoS_2$ L=1/BN L=4, and (c) $WS_2$ L=1/BN L=4/$WS_2$ L=1/BN L=4 heterostructures.

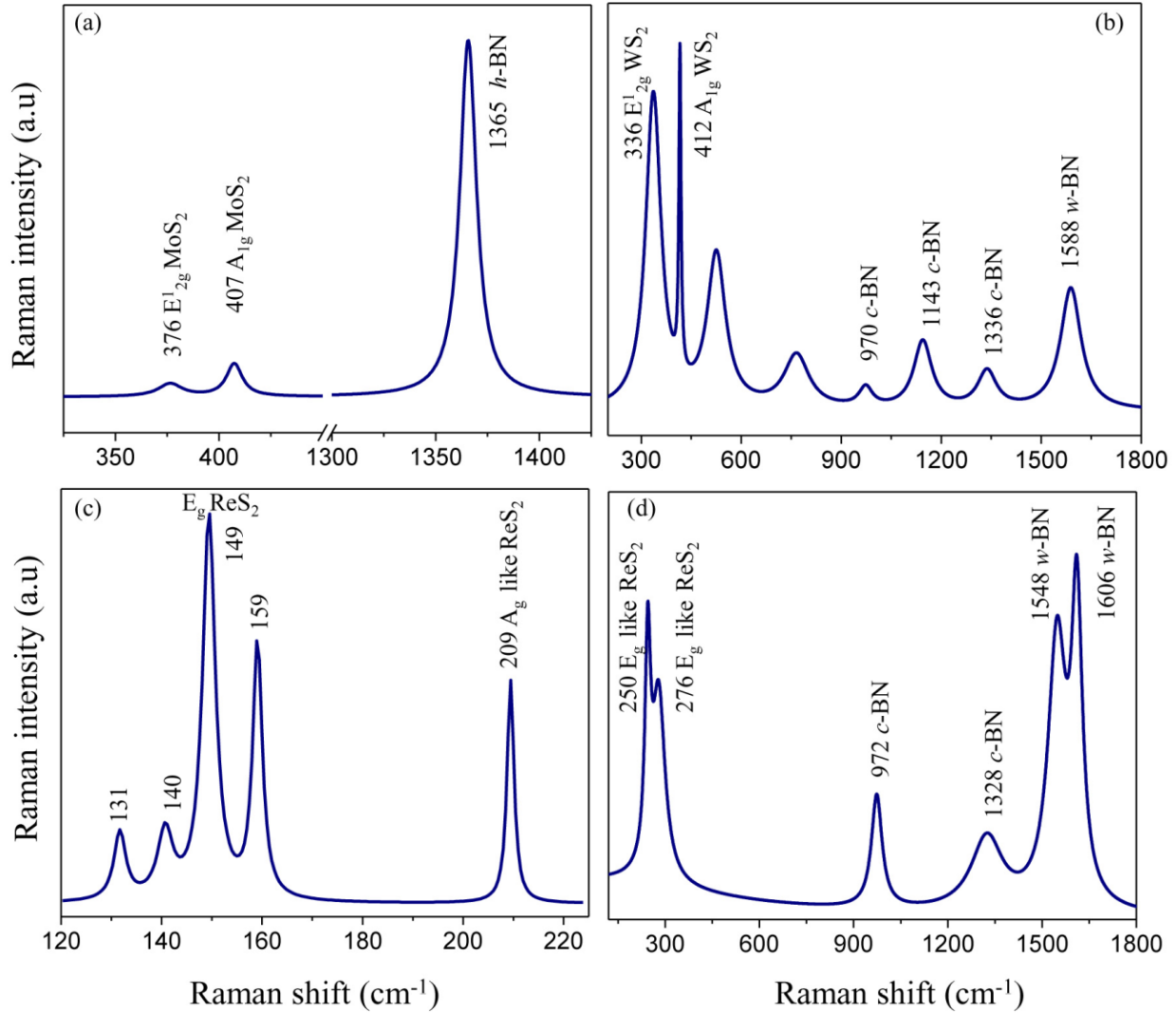

**Figure 2.** Raman spectra of BN grown on top of three different TMDs; (a) on $MoS_2$, (b) $WS_2$ and (c) & (d) $ReS_2$. The structure of BN on top of $MoS_2$ is hexagonal whereas on top of $WS_2$ and $ReS_2$ are mixture of cubic and wurtzite, as indicated in the respective spectra. The Raman spectra for $ReS_2$/BN case are shown in two different panels (c) and (d), respectively.

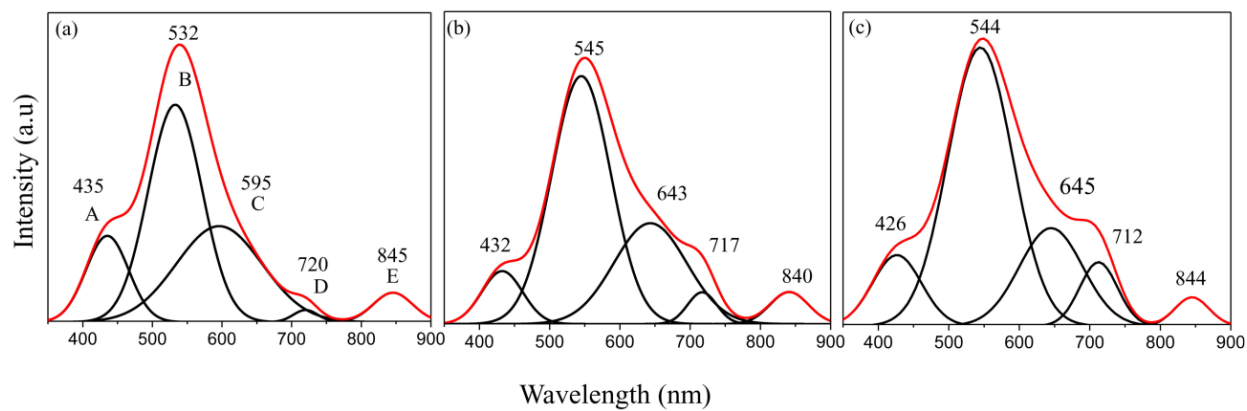

**Figure 3.** Example photoluminescence (PL) spectra of three different heterostructure systems; (a) $MoS_2/BN/MoS_2/BN$, (b) $WS_2/BN/WS_2/BN$, and (c) $ReS_2/BN/ReS_2/BN$. Slight variation is observed in terms of peak positions for different BN interlayers and material systems.

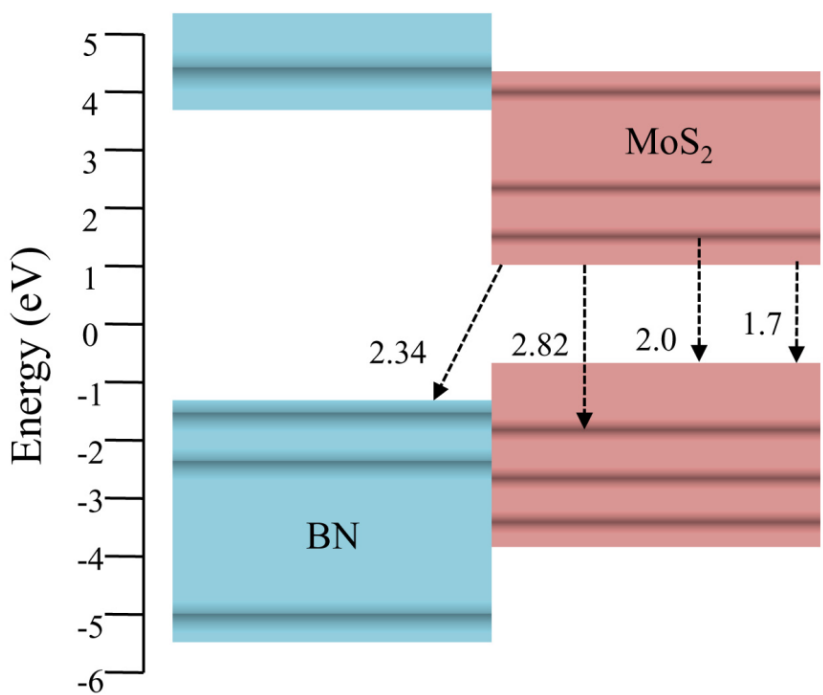

**Figure 4.** Schematic transitions between BN and $MoS_2$ valence and conduction levels [26, 27].

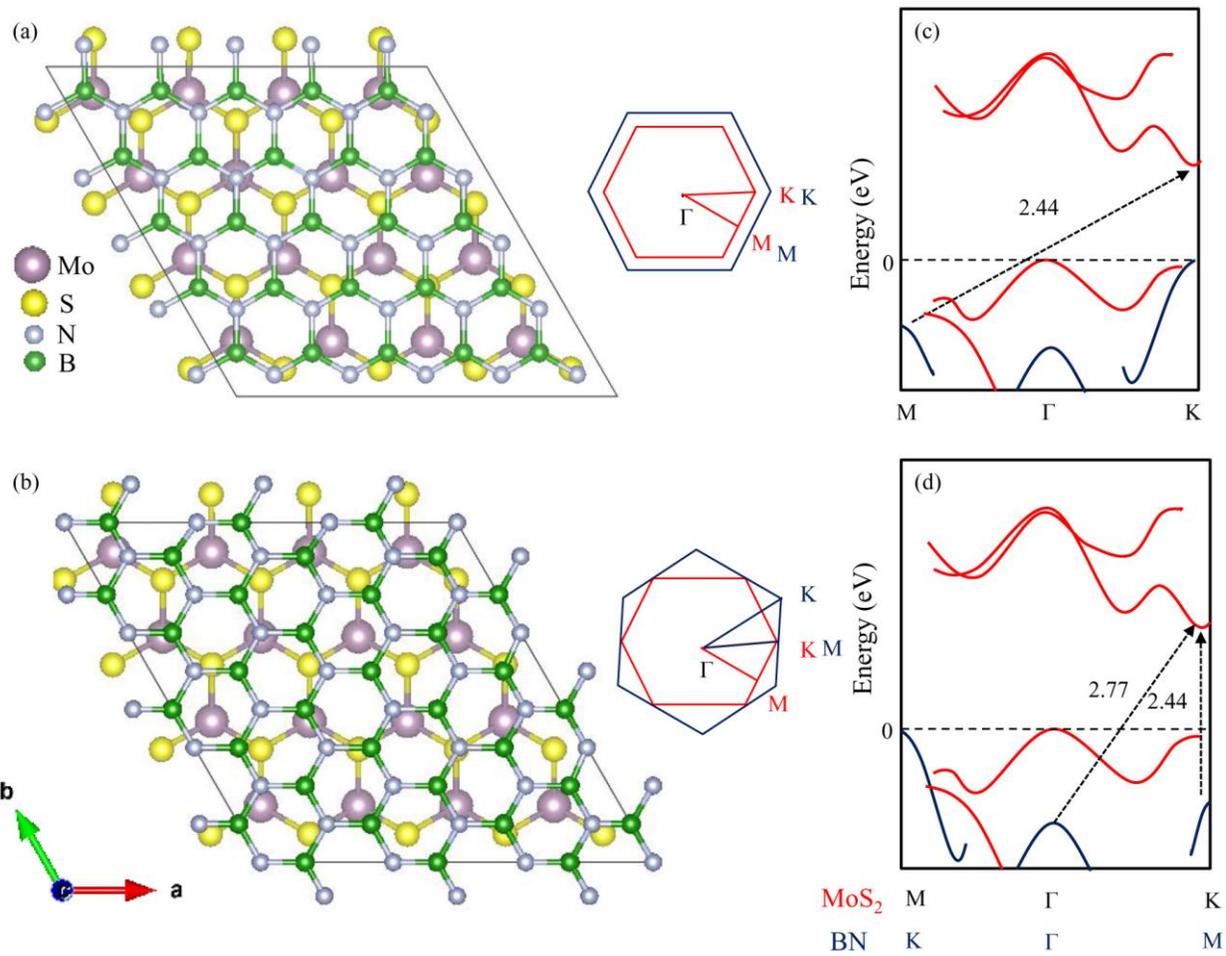

**Figure 5:** Two different *h*-BN stacking on MoS$_2$ (a) Γ → *K* (BN) ∥ Γ → *K*(MoS$_2$) and (b) Γ → *K* (BN) ∥ Γ → *M*(MoS$_2$). Corresponding schematic band structures are shown in (c) and (d), respectively. Kindly note the relative VBM and CBM positions between the two layers with respect to relative rotation between them.

# Supplementary Information

# Distinct photoluminescence in multilayered van der Waals heterostructures of MoS$_2$/WS$_2$/ReS$_2$ and BN

U. Bhat,[1] R. Singh,[1] B. Vishal,[1] Ankit Sharma,[1] H. Sharona,[1] R. Sahu,[1] and R. Datta[1,*]

*[1]International Centre for Materials Science, Chemistry and Physics of Materials Unit, Jawaharlal Nehru Centre for Advanced Scientific Research, Bangalore 560064, India.*

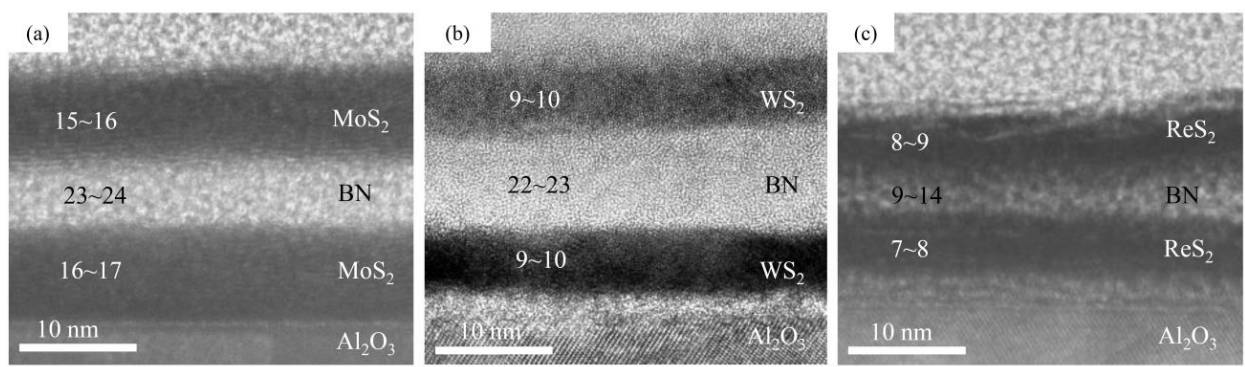

**Figure S1.** Figure S1. Cross sectional TEM images of thicker heterostructures of (a) MoS2/BN/MoS2/BN. (b) WS2/BN/WS2/BN, and (c) ReS2/BN/ReS2/BN. Approximate number of layers are mentioned in the respective images.

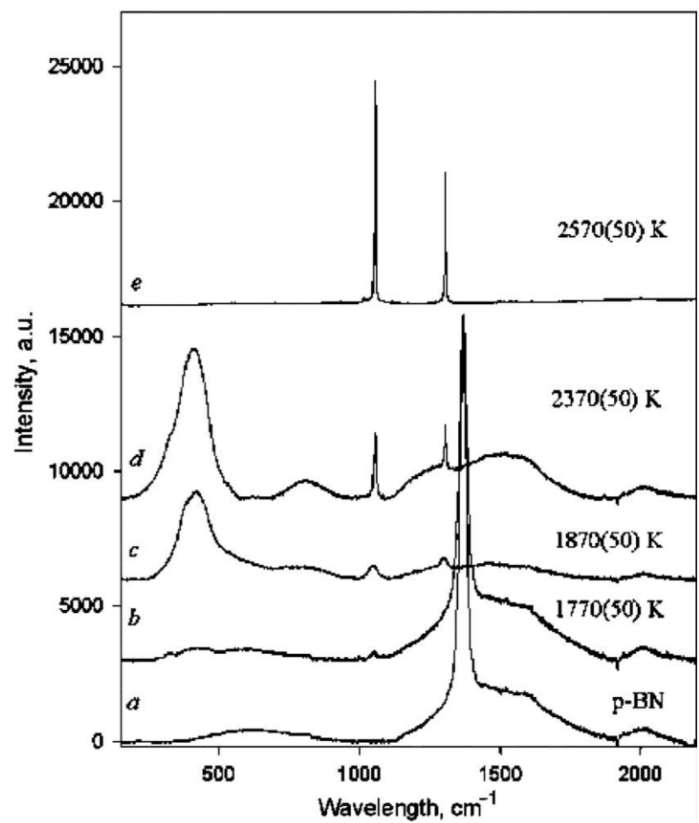

**Figure S2.** Experimental reference Raman spectra for different crystal form of BN. *h*-BN has a unique peak at 1370 cm$^{-1}$, *c*-BN has two sharp peaks at 1057 and 1309 cm$^{-1}$ and *w*-BN has very broad peak around 1600 cm$^{-1}$ wave numbers along with some associated broad peaks [37].

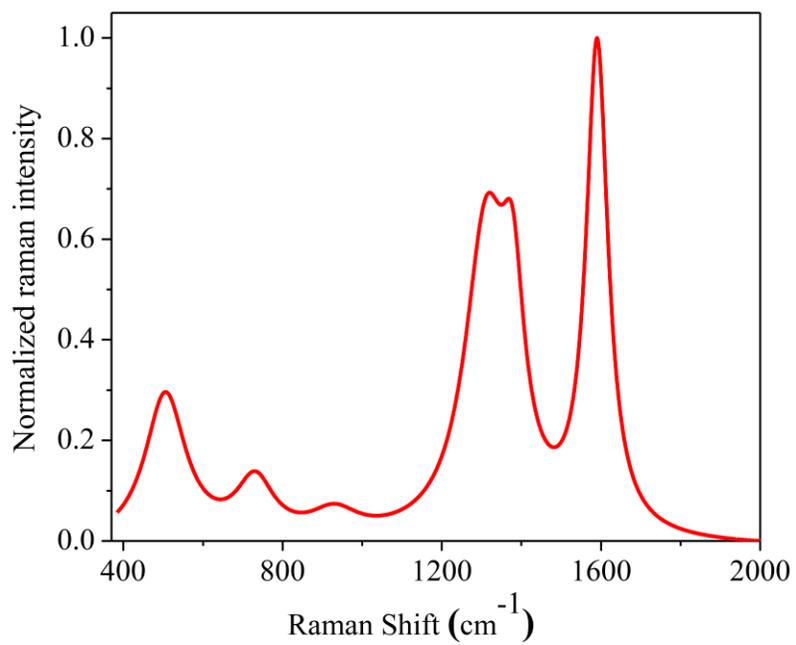

**Figure S3.** Raman spectra of BN grown directly on *c*-plane sapphire confirming the formation of *w*-BN.

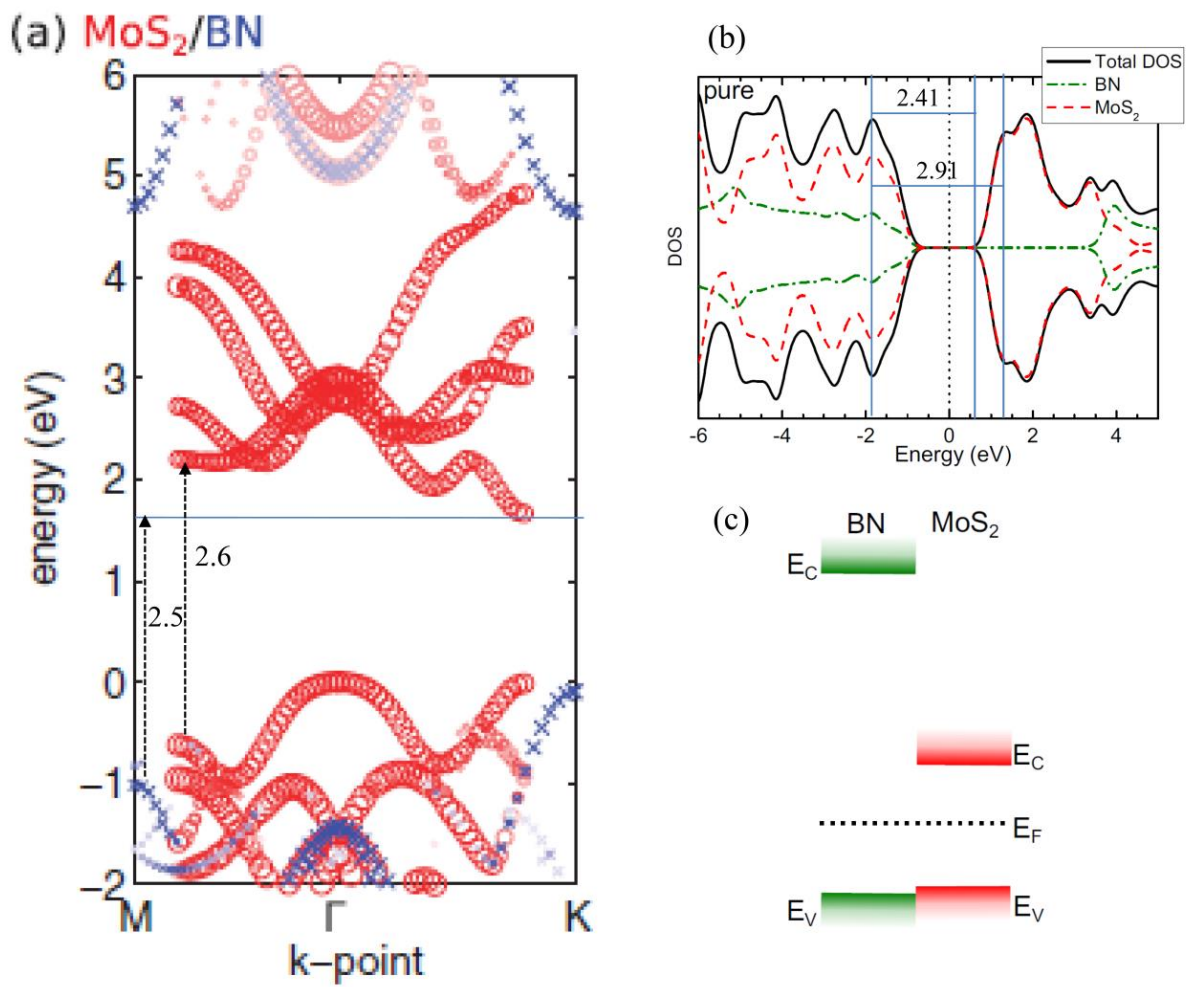

**Figure S4.** Calculated (a) band structure and (b) DOS, and (c) schematic energy band levels of $MoS_2$/BN heterostructure system taken from Ref. 26 & 27. All the possible transitions between BN and $MoS_2$ are marked in the diagram. Calculation predicts a type-I heterostructure formation in this system.

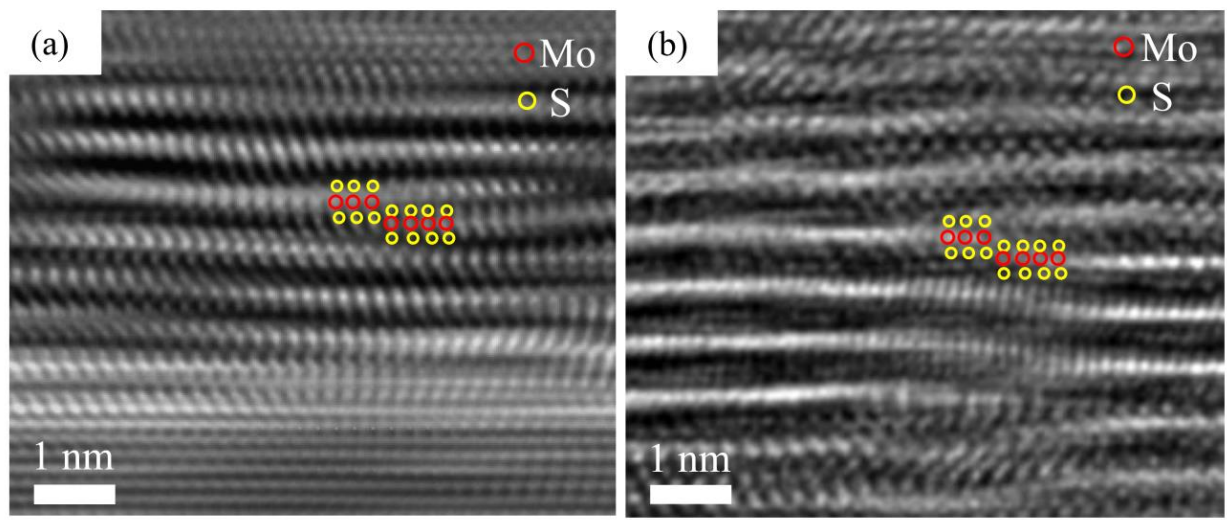

**Figure S5.** HRTEM image of planar fault due to discontinuous Mo atoms in MoS$_2$ epitaxial thin film grown by (a) PLD and (c) ALD. The faults are commonly observed in such TMDs epitaxial thin films.